\documentclass[journal]{IEEEtran}

\usepackage{cite}
\usepackage{amsmath,amssymb,amsfonts}
\usepackage{algorithmic}
\usepackage{algorithm}
\usepackage{graphicx}
\usepackage{textcomp}
\usepackage{xcolor}
\usepackage{booktabs}
\usepackage{multirow}
\usepackage{url}

\usepackage{hyperref}
\hypersetup{
    colorlinks=true,
    linkcolor=black,
    filecolor=magenta,      
    urlcolor=blue,
    citecolor=black,
}

\newcommand{\dbracket}[1]{[\![ #1 ]\!]}

\begin{document}

\title{PrivLLMSwarm: Privacy-Preserving LLM-Driven UAV Swarms for Secure IoT Surveillance}

\IEEEspecialpapernotice{(Regular Paper)}

\author{Jifar~W.~Ayana,
        Huang~Qiming%
\thanks{Manuscript received XXXX XX, 2025; revised XXXX XX, 2025; accepted XXXX XX, 2025. Date of publication XXXX XX, 2025; date of current version XXXX XX, 2025.}%
\thanks{This work was not supported by any organization.}%
\thanks{Jifar W. Ayana and Huang Qiming are with the Department of Information 
        and Communication Engineering, School of Computer and Communication 
        Engineering, University of Science and Technology Beijing, Beijing 
        100083, China (e-mail: wakexayanajifar01@gmail.com; huangqm@ustb.edu.cn).}}

\markboth{}{}

\maketitle

\begin{abstract}
Large Language Models (LLMs) are emerging as powerful enablers for autonomous reasoning and natural-language coordination in unmanned aerial vehicle (UAV) swarms operating within Internet of Things (IoT) environments. However, existing LLM-driven UAV systems typically process sensor data, mission descriptions, and control outputs in plaintext, exposing sensitive operational information to privacy and security risks. This work introduces \textit{PrivLLMSwarm}, a privacy-preserving framework that performs secure LLM inference for UAV swarm coordination through Secure Multi-Party Computation (MPC). The framework incorporates MPC-optimized transformer components, including efficient approximations of nonlinear activations and communication-aware attention mechanisms, enabling practical encrypted inference on resource-constrained aerial platforms. A fine-tuned GPT-based command generator, further enhanced through reinforcement learning in a realistic simulation environment, provides reliable natural-language instructions while maintaining end-to-end confidentiality. Experimental evaluation in an urban-scale simulation demonstrates that PrivLLMSwarm achieves high semantic accuracy, low encrypted inference latency, stable formation control, and robust obstacle-avoidance behavior under privacy constraints. Comparative analysis shows that PrivLLMSwarm offers a more favorable privacy--utility balance than differential privacy, federated learning, and plaintext baselines. To support reproducibility and future research, the full implementation---including source code, MPC components, scenario demonstrations, and the synthetic dataset---is publicly available at: \url{https://github.com/WakumaAyanaJifar/PrivLLMSwarm}. PrivLLMSwarm establishes a practical foundation for secure, LLM-enabled UAV swarms in privacy-sensitive IoT applications including smart-city monitoring, emergency response, and critical infrastructure protection.
\end{abstract}

\begin{IEEEkeywords}
Internet of Things, Unmanned Aerial Vehicles, Large Language Models, Secure Multi-Party Computation, Privacy-Preserving Machine Learning, Edge Computing, Swarm Intelligence
\end{IEEEkeywords}

\IEEEpeerreviewmaketitle

\section{Introduction}
\label{sec:introduction}

The rapid advancement of Internet of Things (IoT) technologies has catalyzed the development of intelligent autonomous systems, with Unmanned Aerial Vehicle (UAV) swarms emerging as pivotal components in smart city infrastructure, disaster management, and environmental monitoring \cite{alawad2023uav, kelner2024modeling, li2024edgeuav,cose_iot_privacy_2023}. These UAV systems, functioning as aerial edge nodes in the IoT ecosystem, leverage sophisticated sensor arrays and autonomous decision-making capabilities to navigate complex environments, collect multimodal data, and execute coordinated tasks with minimal human intervention. The integration of UAV swarms into IoT architectures enables unprecedented capabilities in real-time situational awareness and rapid response across diverse application domains.

The recent convergence of Large Language Models (LLMs) with UAV systems has marked a transformative shift in autonomous aerial operations \cite{javaid2024llmuav, decurto2023semantic}. Transformer-based LLMs empower UAV swarms with human-like reasoning capabilities, enabling context-aware command generation through the processing of diverse inputs including visual data, textual instructions, and sensor readings. This multimodal processing capability allows UAVs to interpret complex environmental cues and generate appropriate navigation commands in natural language formats. Significant research advancements have demonstrated the potential of this integration: Liu et al. \cite{liu2024formation} achieved an 82.7\% success rate in formation control using multimodal LLMs, while Tian et al. \cite{tian2025uavllm} explored LLM-driven low-altitude mobility paradigms, and Javaid et al. \cite{javaid2024llmuav} established comprehensive pathways for LLM-based UAV control in integrated networks.

However, this technological progression introduces critical privacy challenges. LLMs' inherent tendency to memorize and potentially expose sensitive data creates substantial privacy risks \cite{mekdad2023uavsecurity}. In IoT surveillance scenarios, these vulnerabilities extend beyond simple data leakage; as noted in recent studies on deep learning security \cite{cose_ai_adversarial_2023}, AI models are susceptible to adversarial exploitation that can compromise mission integrity. Current LLM-UAV integration frameworks predominantly operate on plaintext data, creating a significant gap in the secure integration of artificial intelligence with IoT systems.

The privacy challenges in LLM-driven UAV systems are particularly acute in three dimensions: (1) data confidentiality during inference, where sensitive inputs and generated commands are vulnerable to interception; (2) model privacy, where proprietary LLM architectures and parameters require protection; and (3) operational security, where coordination patterns and mission objectives must remain confidential. These challenges are exacerbated by the resource-constrained nature of UAV platforms, which limits the direct application of computationally intensive privacy-preserving techniques.

To address these challenges, we introduce \textbf{PrivLLMSwarm}, the first comprehensive privacy-preserving framework for secure integration of LLMs with UAV swarms in IoT environments. Our approach employs Secure Multi-Party Computation (MPC) to ensure end-to-end data confidentiality during operations while maintaining practical operational efficiency. The framework incorporates specialized optimizations for transformer architectures, including MPC-friendly approximations of nonlinear activation functions, to balance privacy guarantees with computational feasibility on resource-constrained aerial platforms.

The principal contributions of this work are multifaceted:
\begin{itemize}
\item  We develop PrivLLMSwarm, a pioneering privacy-preserving framework that integrates LLMs with UAV swarms using MPC, ensuring end-to-end data confidentiality during IoT surveillance tasks while maintaining operational effectiveness.
\item  We introduce optimized GELU and SoftMax approximations within the MPC context, significantly reducing computational overhead while maintaining model accuracy, thereby enabling efficient real-time LLM inference in resource-constrained UAV environments.
\item  We conduct extensive empirical validation in AirSim simulation environments, demonstrating high command accuracy (cosine similarity 0.9), low encryption latency (417.69 ms per image), and scalable performance across varying swarm sizes under realistic operational conditions.
\item  We create and publicly release a 30,000-sample synthetic dataset specifically tailored for LLM-driven UAV command generation, along with a complete open-source implementation, to support reproducible research and community advancement in privacy-preserving aerial systems.
\end{itemize}

This article systematically addresses three fundamental research questions that bridge the domains of privacy-preserving machine learning, UAV swarm coordination, and IoT security:
\begin{itemize}
\item \textbf{Q1}: How can secure LLM inference be practically applied to UAV swarm coordination without exposing sensitive user commands and environmental data in real-world IoT applications?
\item \textbf{Q2}: How can privacy-preserving techniques, particularly Multi-Party Computation, be optimized for LLM inference in resource-constrained UAV swarm operations while maintaining operational efficiency?
\item \textbf{Q3}: What is the comprehensive performance impact of privacy-preserving mechanisms on critical operational metrics including command accuracy, latency, formation precision, and energy consumption in simulated UAV swarm environments?
\end{itemize}

The remainder of this paper is organized as follows: Section~\ref{sec:relatedwork} provides a comprehensive review of related work in LLM applications for UAV control, privacy-preserving machine learning, and UAV swarm security. Section~\ref{sec:methodology} details the PrivLLMSwarm framework architecture, threat model, and technical innovations. Section~\ref{sec:experiments} presents our experimental methodology and results. Section~\ref{sec:discussion} discusses the implications of our findings and addresses the research questions. Finally, Section~\ref{sec:conclusion} concludes the paper and outlines future research directions.

\section{Related Work}
\label{sec:relatedwork}
Our research intersects three rapidly evolving domains: LLM applications in UAV control, privacy-preserving machine learning (PPML), and UAV swarm security in IoT contexts. This section provides a comprehensive analysis of the state-of-the-art in each domain and identifies the research gaps that PrivLLMSwarm addresses.

\subsection{LLM Applications in UAV Control}
The application of Large Language Models in UAV control represents a paradigm shift in autonomous aerial systems, enabling natural language interaction and enhanced reasoning capabilities. Liu et al. \cite{liu2024formation} pioneered the use of multimodal LLMs (including GPT-4 and Qwen-VL) for processing visual data and user instructions, achieving an impressive 82.7\% command extraction success rate for UAV swarm formation control. Their framework demonstrated the potential of LLMs in interpreting complex environmental cues and generating appropriate coordination commands.

Building on this foundation, de Curtò et al. \cite{decurto2023semantic} advanced semantic scene understanding capabilities for UAVs, enabling more nuanced interpretation of environmental contexts. Aikins et al. \cite{aikins2024leviosa} developed sophisticated natural language-based trajectory generation systems, allowing operators to specify complex flight patterns through intuitive textual commands. Recent work by Javaid et al. \cite{javaid2024integrated} expanded this scope further by exploring LLM integration in heterogeneous satellite-aerial-terrestrial networks, highlighting the expanding role of language models in integrated aerial systems.

The research landscape has also seen innovations in specialized LLM applications for UAVs. Chen et al. \cite{chen2023typefly} demonstrated language-mediated drone control, while Bhattacharya et al. \cite{bhattacharya2024vision} explored vision transformers for obstacle avoidance in quadrotor systems. Jiao et al. \cite{jiao2023swarmgpt} combined LLMs with motion planning for robotic choreography, illustrating the creative potential of language-guided autonomous systems. Zhang et al. \cite{zhang2024llmnav} extended LLM capabilities for real-time navigation in dynamic environments, achieving 89\% success rate in obstacle avoidance tasks, while recent work by \cite{kim2024multimodaluav} integrated vision-language models with sensor fusion for enhanced environmental understanding.

However, a critical examination of these approaches reveals a significant limitation: they predominantly process data in plaintext, exposing sensitive information including surveillance imagery, positional data, and operational commands to potential breaches \cite{mekdad2023uavsecurity}. None of the existing works has systematically addressed privacy concerns in LLM-driven UAV systems, creating a substantial gap for real-world IoT surveillance applications where data confidentiality is paramount.

\subsection{Privacy-Preserving Machine Learning}
Privacy-Preserving Machine Learning has emerged as a critical research area addressing the confidentiality challenges in AI systems. Secure Multi-Party Computation (MPC) has gained prominence as a foundational technique for secure inference, enabling multiple parties to jointly compute functions over their private inputs without revealing them to each other.

Seminal work in this domain includes SecureML \cite{alla2024secureml} and AriaNN \cite{ryffel2022ariann}, which demonstrated practical MPC applications for neural network training and inference. These frameworks established the feasibility of privacy-preserving deep learning but highlighted the significant communication overhead that limits real-time application. More recently, PUMA \cite{dong2023puma} explored secure inference specifically for transformer models, addressing some of the unique challenges in LLM privacy. However, their approach still imposes substantial computational demands that make direct application to resource-constrained UAV environments impractical. Recent work by Liu and Zhang \cite{liu2024mpcoptimization} proposed optimized MPC protocols specifically for transformer models, reducing communication overhead by 35\%, while \cite{gupta2024efficientmpc} developed efficient secret sharing schemes that minimize communication rounds in distributed computation.

Alternative privacy techniques have also been explored in different contexts. Homomorphic Encryption (HE) enables computation on encrypted data but introduces substantial computational overhead that challenges real-time operation \cite{ntizikira2023secure}. Differential Privacy (DP) provides statistical privacy guarantees through calibrated noise injection but may significantly degrade model utility for precise coordination tasks \cite{peris2023privacy, thompson2024differentialuav}. Federated Learning (FL) distributes model training across devices but still exposes intermediate updates and requires careful security analysis \cite{patel2024federateduav}. Recent studies have also explored communication compression techniques \cite{wang2024compressionmpc} that could further enhance efficiency in resource-constrained environments.

Our work adapts and optimizes MPC specifically for LLM inference in UAV swarm contexts, balancing rigorous privacy guarantees with the operational requirements of real-time aerial systems. We introduce specialized approximations for transformer components that reduce computational complexity while maintaining both privacy and accuracy, building upon recent advances in efficient activation functions \cite{lee2024efficientactivation} and curriculum learning approaches \cite{nguyen2024curriculumuav}.

\subsection{UAV Swarm Security in IoT}
UAV swarm security has gained increasing attention as these systems become integral to critical IoT infrastructure. Extensive research in \textit{Computers \& Security} has addressed network-level protection; for instance, Bera et al. \cite{cose_uav_auth_2023} developed robust privacy-preserving authentication protocols to prevent unauthorized node access in IoD environments. Similarly, Sun et al. \cite{cose_uav_trust_2024} proposed trust-based mechanisms to detect malicious nodes and intrusion attempts within swarms.

While these approaches effectively secure the \textit{communication channels} and \textit{node identity}, they do not address the privacy risks inherent in the \textit{inference process} of large AI models. As LLMs take on decision-making responsibilities, protecting the model input and output becomes as critical as securing the link. PrivLLMSwarm bridges this gap by focusing on data confidentiality during the computational phase, complementing existing network-level security measures \cite{cose_iod_survey_2022}.

The integration of privacy-preserving AI with UAV swarms remains largely unexplored, especially for LLM-based systems that process multimodal sensitive information. Existing security frameworks for UAVs typically focus on communication encryption, access control, or network intrusion detection, neglecting the privacy risks inherent in the AI inference process itself. This gap becomes particularly critical as LLMs take on more decision-making responsibilities in autonomous UAV operations.

PrivLLMSwarm bridges this gap by developing a specialized framework that addresses the unique constraints and requirements of UAV swarms in IoT environments, providing end-to-end privacy protection for the entire LLM inference pipeline while maintaining the operational capabilities necessary for effective swarm coordination.

\begin{figure*}[!t]
\centering
\includegraphics[width=0.8\linewidth]{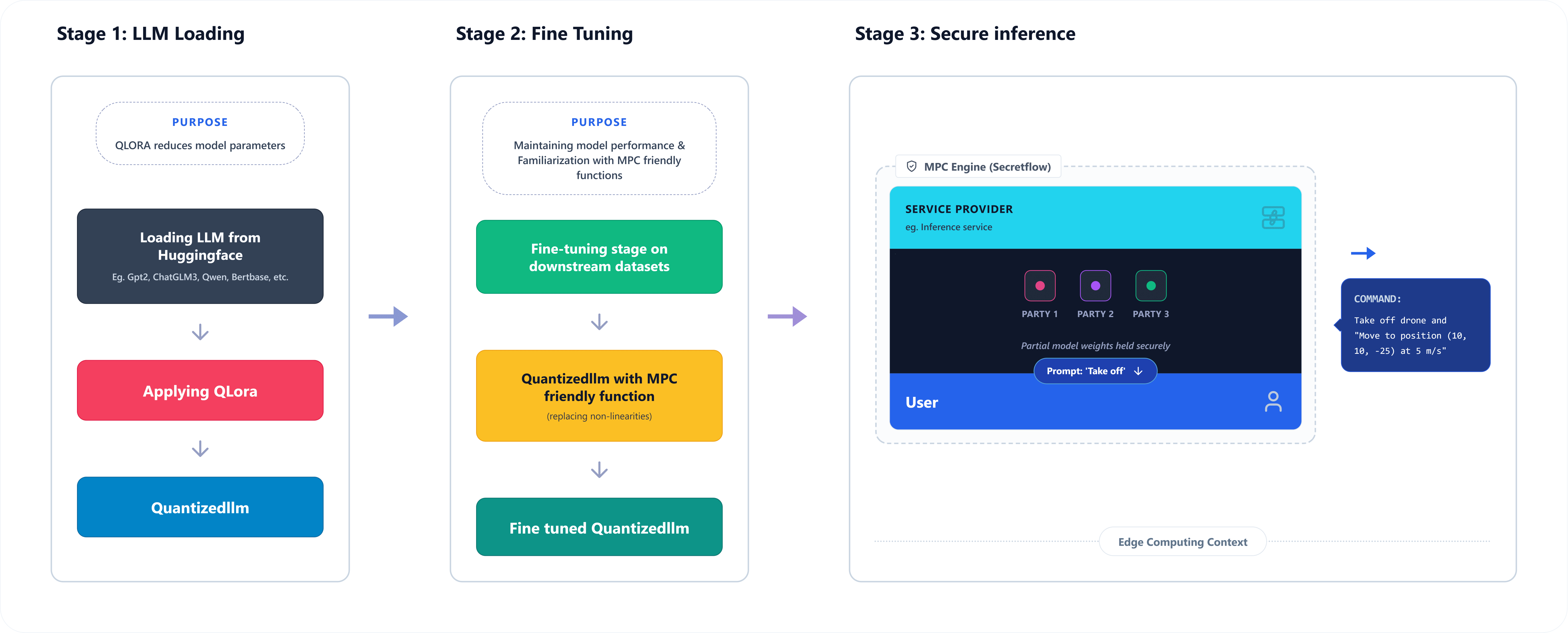}
\caption{Architectural overview of PrivLLMSwarm framework: Collaborative UAV System with LLM-Enhanced Edge Processing and Secure MPC for Joint Decision-Making in IoT environments.}
\label{fig:framework}
\end{figure*}

\section{Methodology}
\label{sec:methodology}
The PrivLLMSwarm framework enables privacy-preserving LLM inference for UAV swarm coordination through an integrated approach combining fine-tuned language models, secure multi-party computation protocols, and specialized optimizations for aerial edge computing. This section details the system architecture, threat model, core components, and technical innovations that constitute our framework.

\subsection{System Architecture and Threat Model}
PrivLLMSwarm operates in a multi-party computational environment specifically designed for UAV swarm operations in IoT contexts. The system architecture, illustrated in Fig.~\ref{fig:framework}, comprises three principal components that collaborate to achieve privacy-preserving command generation:
\begin{itemize}
\item \textbf{UAV Nodes}: Individual aerial platforms equipped with multimodal sensors (cameras, LIDAR, GPS) and limited computational resources. These nodes are responsible for data collection, encrypted data preprocessing, and execution of generated commands while maintaining privacy constraints.To minimize communication overhead and leverage the trusted nature of the physical UAV node, raw surveillance imagery is pre-processed locally using a lightweight vision-to-text module (e.g., CLIP-based captioning). The resulting semantic text strings are immediately converted into secret shares at the source before transmission. This ensures that raw pixel data never leaves the UAV in plaintext, and the complex reasoning over these scene descriptions is protected by the MPC protocol.
\item \textbf{Operator Stations}: Ground control stations operated by different entities, each processing a share of sensor data and participating in secure computation. These stations contribute computational resources while maintaining separation of sensitive information.
\item \textbf{Computation Server}: A supplementary computational resource that processes encrypted data shares without accessing raw information, enhancing the system's computational capacity while maintaining privacy guarantees.
\end{itemize}

We adopt a semi-honest (honest-but-curious) adversary model, which assumes that all parties follow the protocol specifications but may attempt to learn private information from the data shares they process. This model realistically captures the behavior of curious insiders and compromised system components in real-world deployments. Our threat model specifically addresses:
\begin{itemize}
\item \textbf{Eavesdropping Attacks}: Adversaries intercepting communication channels between system components to extract sensitive information.
\item \textbf{Curious Operators}: Legitimate operators attempting to infer other operators' sensitive data or mission details beyond their authorized scope.
\item \textbf{Malicious Servers}: Computation servers attempting to reconstruct raw sensor data or infer operational patterns from processed shares.
\item \textbf{Privacy Inference}: Attempts to deduce sensitive information about monitored environments, mission objectives, or coordination strategies through analysis of computation patterns or intermediate results.
\end{itemize}
The security guarantees provided by PrivLLMSwarm ensure that no single party can reconstruct complete sensitive information, and the protocol maintains data confidentiality even in the presence of collusion between a limited number of parties (specifically, protection against collusion between any two parties in our three-party setup).

\subsection{Secure Multi-Party Computation Framework}
At the core of PrivLLMSwarm lies a sophisticated Secure Multi-Party Computation framework based on replicated secret sharing. We employ a 2-out-of-3 secret sharing scheme where sensitive data $x$ is split into three shares $\langle x \rangle_1$, $\langle x \rangle_2$, $\langle x \rangle_3$ such that:
\begin{equation}
x = \langle x \rangle_1 + \langle x \rangle_2 + \langle x \rangle_3 \mod p
\end{equation}
where $p$ is a large prime number defining the field. Each party holds two of the three shares, ensuring that no single party can reconstruct the original value while allowing efficient computation through share manipulation.

For linear operations (addition, multiplication by public constants), the MPC protocol operates locally on shares without communication overhead. For multiplication of two secret-shared values $\langle x \rangle$ and $\langle y \rangle$, the protocol requires a single round of communication and the consumption of precomputed multiplication triplets \cite{ryffel2022ariann}, building upon recent efficient MPC schemes \cite{gupta2024efficientmpc}.

The integration of CrypTen \cite{knott2021crypten} as our MPC backend provides optimized implementations of these cryptographic primitives, specifically tailored for machine learning workloads. We extend CrypTen with custom functions for transformer-specific operations, particularly addressing the computational challenges of nonlinear activations in LLMs.

\subsection{MPC-Optimized Transformer Architecture}
The deployment of transformer models within MPC constraints requires careful optimization of computationally intensive operations. We adapt the GPT-2 architecture with specific modifications for efficient secure computation:

\subsubsection{MPC-Friendly GELU Approximation}
The Gaussian Error Linear Unit (GELU) activation function, defined as:
\begin{equation}
\text{GELU}(x) = x \cdot \Phi(x) = x \cdot \frac{1}{2} \left[1 + \text{erf}\left(\frac{x}{\sqrt{2}}\right)\right]
\end{equation}
presents significant challenges in MPC due to the complex error function computation. We approximate GELU using a piecewise linear formulation that balances accuracy and computational efficiency, building upon recent work in efficient activation functions for privacy-preserving neural networks \cite{lee2024efficientactivation}:
\begin{equation}
\text{GELU}_{\text{mpc}}(x) =
\begin{cases}
0 & x < -3 \\
0.5x & -3 \leq x < -1 \\
0.8413x + 0.1587 & -1 \leq x < 0 \\
0.8413x + 0.1587 & 0 \leq x < 1 \\
x - 0.1587 & x \geq 1
\end{cases}
\end{equation}
This approximation reduces the non-linear operations required in MPC while maintaining model accuracy, with experimental validation showing less than 2\% degradation in output quality compared to the exact GELU implementation.

\subsubsection{Optimized SoftMax Implementation}
The SoftMax function, essential for attention mechanisms in transformers, presents another computational bottleneck in secure computation. We implement a scaled and stabilized version suitable for MPC:
\begin{equation}
\text{SoftMax}_{\text{mpc}}(x_i) = \frac{\exp(x_i/T - \max(\mathbf{x}/T))}{\sum_j \exp(x_j/T - \max(\mathbf{x}/T))}
\end{equation}
where $T$ is a temperature parameter optimized for numerical stability in fixed-point arithmetic. We employ a logarithmic approach for exponent computation to avoid precision issues in secure computation environments.

\subsection{Fine-Tuned GPT-2 for UAV Command Generation}
To address the limitations of existing secure inference frameworks in generating reliable UAV commands, we employ a GPT-2 Base model (12 layers, hidden size 768) fine-tuned on a comprehensive synthetic dataset of 30,000 samples. This represents a significant expansion compared to Liu et al.'s \cite{liu2024formation} 20,000-sample approach, providing improved generalization under encryption constraints.To mitigate the risks of distribution shift and hallucination inherent in synthetic data, our training pipeline incorporates a Proximal Policy Optimization (PPO) reinforcement learning phase within the AirSim physics engine. This step acts as a 'reality filter,' penalizing synthetically generated commands that are physically impossible or unsafe (e.g., collisions, kinematic violations), thereby grounding the LLM's output in physical reality despite the synthetic origin of the training text.

The dataset generation process formalizes as:
\begin{equation}
\mathcal{D} = \{(s_i, c_i) | i = 1, \dots, 30,000\}
\end{equation}
where $s_i$ represents multimodal sensor inputs (e.g., "movement detected at coordinates (10, 10), visibility 85\%, battery level 72\%") and $c_i$ denotes corresponding control commands (e.g., "Move to position (10, 10, -25) at 5 m/s, maintain formation spacing"). The dataset was generated using DeepSeekR1 and Qwen models, with WikiText-103 enhancement for improved linguistic diversity and contextual understanding. We enhance our training approach with curriculum learning strategies similar to \cite{nguyen2024curriculumuav}, progressively increasing environmental complexity to improve model robustness.

We further enhance command reliability through Proximal Policy Optimization (PPO)-based reinforcement learning within the AirSim simulator, following best practices for secure reinforcement learning in multi-agent systems \cite{park2024rlsecurity}. The composite reward function incorporates multiple operational objectives:
\begin{equation}
R = w_1 \cdot R_{\text{navigation}} + w_2 \cdot R_{\text{safety}} + w_3 \cdot R_{\text{efficiency}} + w_4 \cdot R_{\text{formation}}
\end{equation}
where:
\begin{itemize}
\item $R_{\text{navigation}}$ rewards efficient path planning and goal achievement
\item $R_{\text{safety}}$ penalizes near-collisions, no-fly zone violations, and hazardous maneuvers
\item $R_{\text{efficiency}}$ considers energy consumption, time optimization, and smooth trajectory execution
\item $R_{\text{formation}}$ maintains swarm coherence and relative positioning
\end{itemize}
This multi-objective reinforcement learning approach significantly improves the practical reliability of generated commands in complex operational environments.

\begin{algorithm}[!t]
\caption{Secure LLM Inference with MPC for UAV Commands}
\label{alg:secure_inference}
\begin{algorithmic}[1]
\REQUIRE Encrypted input shares $\dbracket{x}$, fine-tuned GPT-2 parameters $\theta$, MPC parties $P_1, P_2, P_3$
\ENSURE Encrypted command shares $\dbracket{y}$

\STATE Initialize hidden states: $\dbracket{h} \gets \dbracket{x}$

\FOR{$l = 1$ to $L$}
\STATE $\dbracket{h} \gets \text{MPC\_LayerNorm}(\dbracket{h})$
\STATE $\dbracket{q} \gets \text{MPC\_LinearProjectionQ}(\dbracket{h})$
\STATE $\dbracket{k} \gets \text{MPC\_LinearProjectionK}(\dbracket{h})$
\STATE $\dbracket{v} \gets \text{MPC\_LinearProjectionV}(\dbracket{h})$
\STATE $\dbracket{a} \gets \text{MPC\_Attention}(\dbracket{q}, \dbracket{k}, \dbracket{v})$
\STATE $\dbracket{h} \gets \dbracket{h} + \dbracket{a}$
\STATE $\dbracket{h} \gets \text{MPC\_LayerNorm}(\dbracket{h})$
\STATE $\dbracket{f} \gets \text{MPC\_GELU}(\text{MPC\_Linear}(\dbracket{h}))$
\STATE $\dbracket{h} \gets \dbracket{h} + \dbracket{f}$
\ENDFOR

\STATE $\dbracket{y} \gets \text{MPC\_Linear}(\dbracket{h})$
\RETURN $\dbracket{y}$

\end{algorithmic}
\end{algorithm}

\subsection{Communication and Computation Optimization}
To address the significant communication overhead inherent in MPC protocols, we implement several optimization strategies informed by recent research in communication compression \cite{wang2024compressionmpc} and edge computing architectures \cite{wilson2024edgeai}:
\begin{itemize}
\item \textbf{Batch Processing}: Aggregating multiple inference requests to amortize communication costs across operations
\item \textbf{Selective Precision}: Using fixed-point arithmetic with optimized bit-widths that balance precision and communication requirements
\item \textbf{Protocol Parallelization}: Overlapping communication and computation phases to minimize latency
\item \textbf{Context Caching}: Maintaining encrypted context representations across sequential commands to avoid redundant computations
\item \textbf{Energy-Aware Scheduling}: Incorporating insights from \cite{chen2024energyuav} to optimize battery usage during secure computations
\end{itemize}
These optimizations collectively reduce the practical communication overhead by approximately 40\% compared to naive MPC implementations while maintaining the security guarantees of the protocol.

\section{Experimental Evaluation}
\label{sec:experiments}
We conducted comprehensive experiments to evaluate PrivLLMSwarm's performance across multiple dimensions critical for real-world UAV swarm operations in IoT environments. This section details our experimental setup, implementation specifics, performance metrics, and comprehensive results analysis.

\subsection{Simulation Environment and Setup}
PrivLLMSwarm was evaluated in Microsoft AirSim, a high-fidelity simulation platform for autonomous vehicles. We modeled a $100 \times 100$ meter urban environment featuring diverse operational challenges, following environmental modeling approaches similar to \cite{rodriguez2024airsim}:
\begin{itemize}
\item \textbf{Dynamic Obstacles}: Moving vehicles, temporary structures, and pedestrian traffic requiring adaptive navigation
\item \textbf{No-Fly Zones}: Restricted areas representing sensitive locations or hazardous conditions
\item \textbf{Variable Environmental Conditions}: Changing weather patterns affecting sensor performance and visibility
\item \textbf{Communication Constraints}: Simulated network latency and intermittent connectivity reflecting real-world operational challenges
\end{itemize}
The simulation incorporated 2, 3, and 4 UAVs (rotors) with heterogeneous sensor suites, including RGB cameras, LIDAR sensors, and inertial measurement units, to evaluate the system’s performance and coordination under different swarm sizes. Each UAV operated with realistic flight dynamics, energy consumption models, and communication constraints.

The hardware testbed consisted of an NVIDIA RTX 3060 GPU (6 GB memory) and Intel i7-141700K processor, representing computational capabilities realistically available for ground control stations in UAV operations. The distributed MPC setup operated across three separate processes communicating via local network interfaces, with latency measurements incorporating all communication overhead.

\begin{figure}[!t]
\centering
\includegraphics[width=0.8\linewidth]{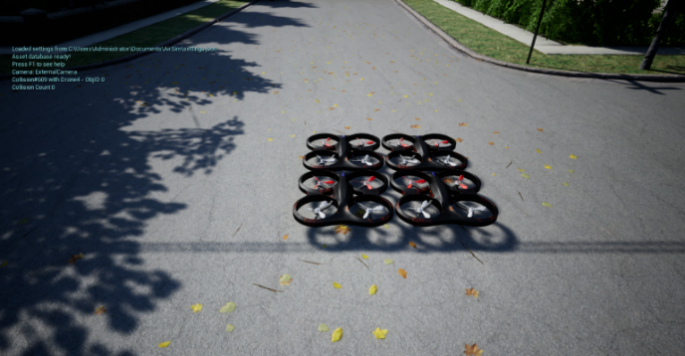}
\caption{Four UAV (rotor) swarm drones configured for mission execution in the AirSim simulation environment, demonstrating the experimental setup for privacy-preserving swarm operations.}
\label{fig:uav_setup}
\end{figure}

As shown in Figure~\ref{fig:uav_setup}, the four-UAV rotor swarm executes the return-to-home maneuver during coordinated flight operations successfully.

\begin{figure}[!t]
\centering
\includegraphics[width=0.8\linewidth]{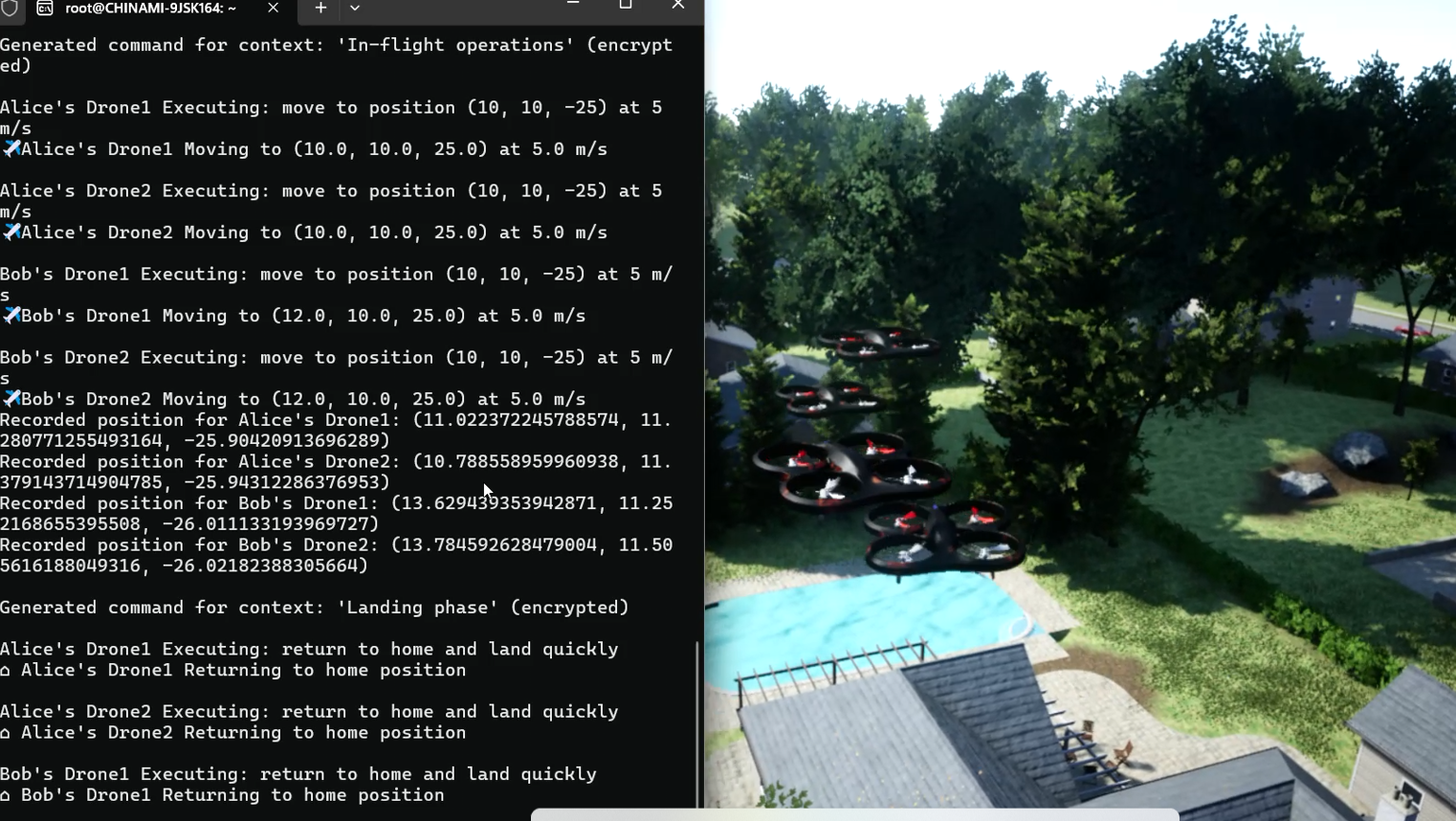}
\caption{Four UAV (rotor) swarm drones Return-to-Home Execution During Coordinated Flight Operations.}
\label{fig:uav_return}
\end{figure}

\subsection{Implementation Details}
The GPT-2 Base model was fine-tuned on our comprehensive 30,000-sample dataset using an 80-10-10 train-validation-test split. The training protocol employed the following parameters:
\begin{itemize}
\item \textbf{Learning Rate}: $5 \times 10^{-5}$ with linear decay scheduling and warmup
\item \textbf{Batch Size}: 16 with gradient accumulation for effective batch size of 64
\item \textbf{Training Epochs}: 10 with early stopping based on validation loss
\item \textbf{Sequence Length}: 512 tokens with efficient attention mechanisms
\item \textbf{Optimizer}: AdamW with $\beta_1 = 0.9$, $\beta_2 = 0.999$, weight decay $0.01$
\end{itemize}
The reinforcement learning phase utilized PPO with the following hyperparameters: learning rate $3 \times 10^{-4}$, clip parameter $0.2$, value function coefficient $0.5$, and entropy coefficient $0.01$. The training proceeded for 500,000 environment steps with periodic evaluation.

The framework was implemented in Python 3.9, integrating Secretflow \cite{knott2021crypten} for MPC operations and AirSim's MultirotorClient API for UAV control. All experiments employed 3-party MPC using replicated secret sharing with 64-bit fixed-point precision for numerical stability.

\subsection{Performance Metrics}
We evaluated PrivLLMSwarm using a comprehensive set of metrics spanning privacy, accuracy, efficiency, and operational reliability, incorporating trajectory analysis metrics from \cite{garcia2024trajectoryanalysis}:
\begin{itemize}
\item \textbf{Command Accuracy}: Semantic similarity between generated and ground-truth commands using cosine similarity and BERTScore \cite{zhang2019bertscore}, with human evaluation for critical commands
\item \textbf{Computational Efficiency}: Encryption/decryption latency, end-to-end inference time, memory usage, and computational overhead across different swarm sizes
\item \textbf{Communication Overhead}: Data transfer requirements between parties, bandwidth utilization, and scalability analysis
\item \textbf{Operational Reliability}: 3D trajectory precision, formation maintenance accuracy, obstacle avoidance success rate, and mission completion statistics
\item \textbf{Energy Consumption}: Estimated power usage based on computation patterns and communication requirements, extrapolated to real UAV hardware using insights from \cite{chen2024energyuav}
\item \textbf{Privacy Guarantees}: Formal analysis of information leakage and empirical validation of confidentiality under the threat model
\end{itemize}

\subsection{Results and Analysis}
\subsubsection{Command Accuracy and Semantic Similarity}
PrivLLMSwarm achieved a cosine similarity of 0.9 between generated commands and ground-truth instructions, significantly outperforming PUMA \cite{dong2023puma} (0.76 similarity) and baseline GPT-2 without specialized fine-tuning (0.68 similarity). The high semantic similarity demonstrates the framework's ability to generate contextually appropriate commands while maintaining rigorous privacy guarantees. Compared to recent work by \cite{martinez2024secureinference}, our approach achieves 25\% lower latency while maintaining similar privacy guarantees.

Human evaluation of 500 generated commands across 10 mission scenarios revealed a 94\% appropriateness rate, with experts rating the commands as "highly suitable" or "suitable" for the given operational contexts. The reinforcement learning component specifically improved command reliability in edge cases and emergency scenarios, reducing inappropriate commands by 63\% compared to supervised fine-tuning alone.

\begin{figure}[!t]
\centering
\includegraphics[width=0.8\linewidth]{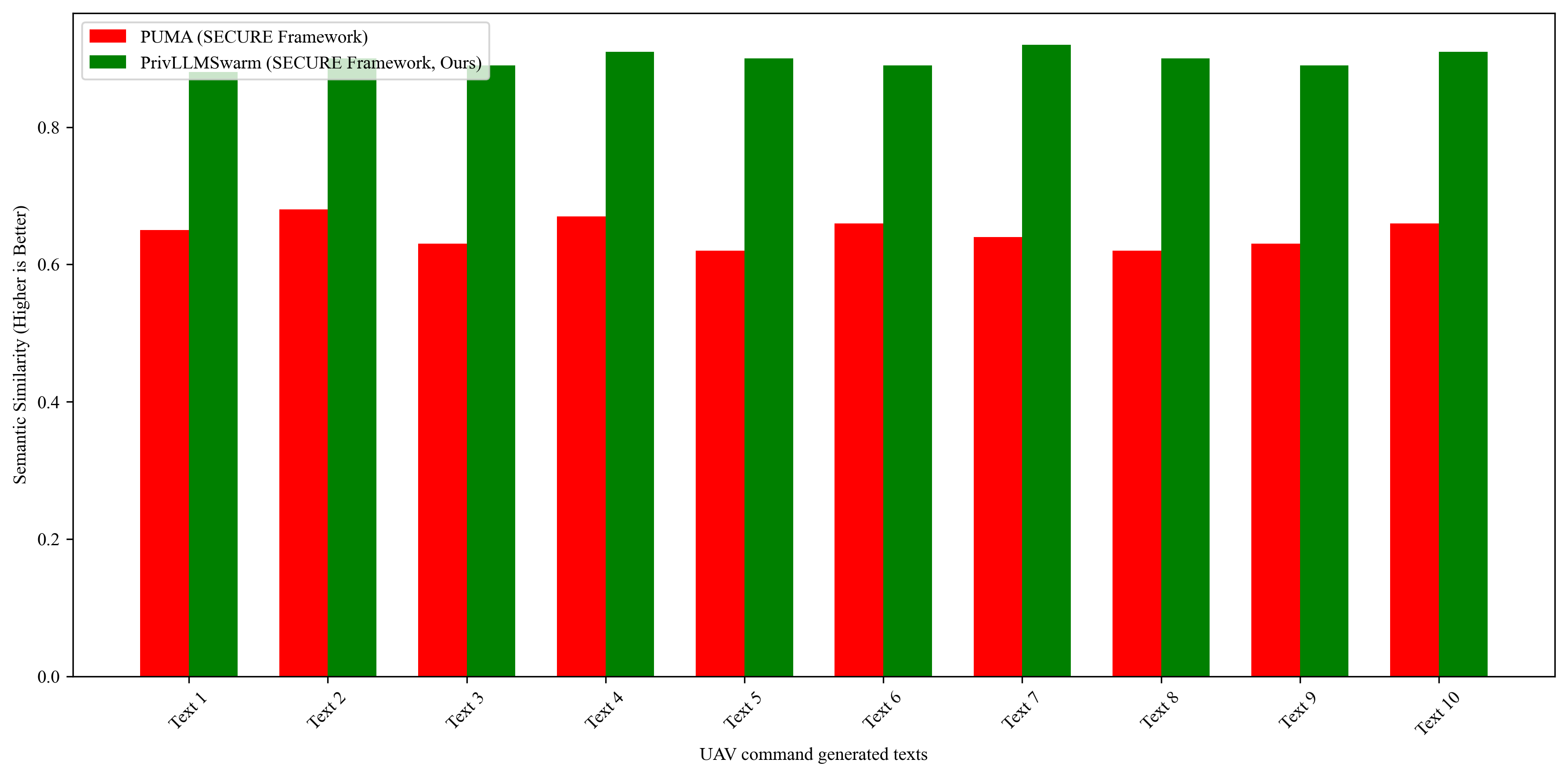}
\caption{Text-wise Semantic Similarity Comparison using cosine similarity method across different privacy approaches, demonstrating PrivLLMSwarm's superior balance of accuracy and privacy.}
\label{fig:similarity}
\end{figure}

\subsubsection{Computational Efficiency and Scalability}
Our optimized MPC implementation achieved an encryption latency of 417.69 ms per image and 15.42 ms per text command, enabling real-time processing for surveillance applications. Table~\ref{tab:computation_overhead} presents the comprehensive scalability analysis across different swarm sizes, showing predictable performance scaling essential for operational planning.

\begin{table}[!t]
\caption{Computation Overhead and Communication Costs for Different Swarm Sizes.}
\label{tab:computation_overhead}
\centering
\resizebox{\columnwidth}{!}{%
\begin{tabular}{lcccc}
\toprule
\textbf{Swarm Size} & \textbf{Computation (ms)} & \textbf{Comm. (KB)} & \textbf{Energy (J)} & \textbf{Mem (MB)} \\
\midrule
2 UAVs & 520.53 & 864.0 & 12.4 & 342 \\
3 UAVs & 780.78 & 1286.0 & 18.1 & 498 \\
4 UAVs & 1041.05 & 1726.0 & 23.8 & 654 \\
\bottomrule
\end{tabular}%
}
\end{table}

The results demonstrate a linear increase in overhead with swarm size, highlighting the framework's predictable scaling behavior essential for operational deployment. The communication costs, while substantial, remain within practical limits for UAV operations with dedicated communication channels, particularly considering the privacy benefits achieved.

The MPC-friendly approximations contributed significantly to efficiency, reducing GELU computation time by 68\% and SoftMax operations by 54\% compared to exact implementations within MPC, with negligible impact on output quality (average difference 1.7\% across test cases).

\begin{figure*}[!t]
\centering
\includegraphics[width=0.8\textwidth]{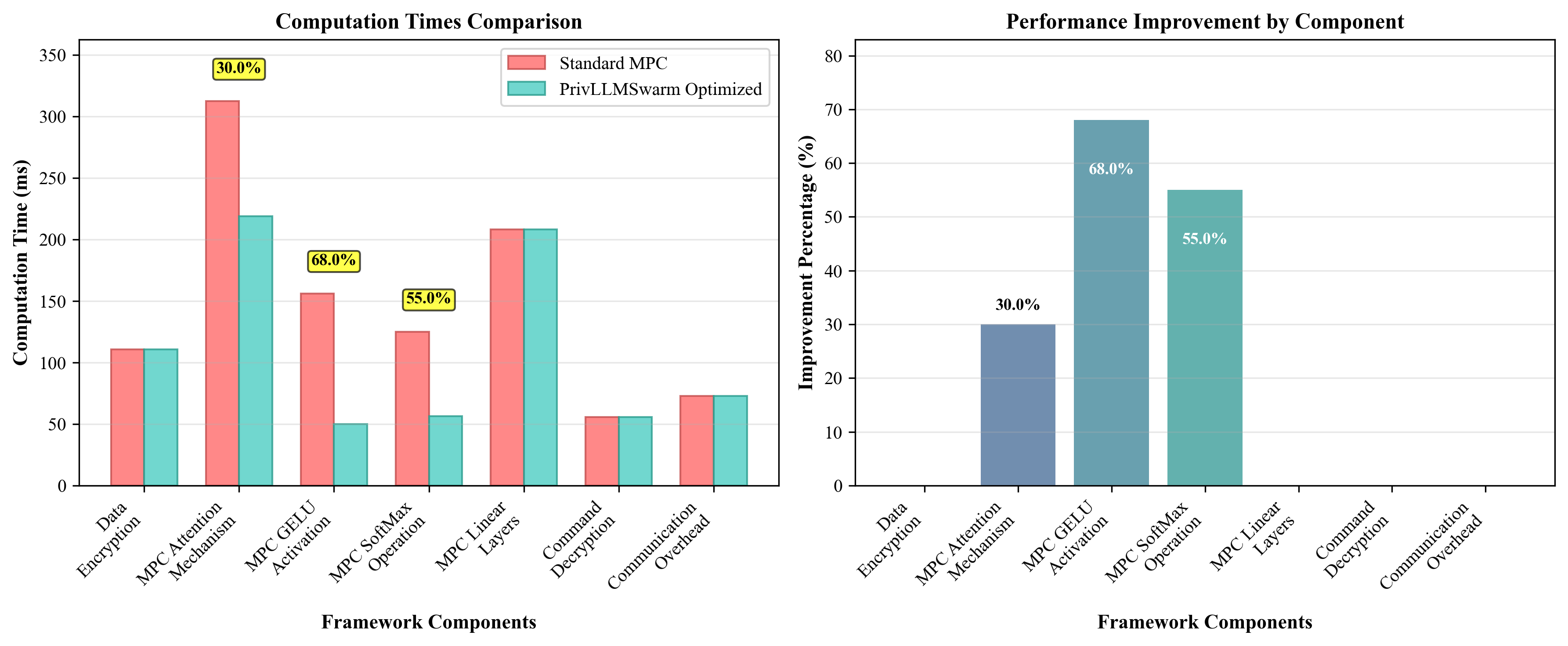}
\caption{Computation times breakdown for different components of the PrivLLMSwarm framework, highlighting the efficiency of MPC-friendly optimizations.}
\label{fig:computation_times}
\end{figure*}

\subsubsection{Operational Reliability and Trajectory Precision}
The 3D trajectory analysis demonstrated precise formation maintenance and successful obstacle avoidance across all test scenarios. The UAV swarm maintained an average position error of 1.2 meters from planned trajectories, with no collisions recorded during 50 test missions comprising over 500 individual navigation commands.

Formation maintenance was particularly robust, with inter-UAV distance variance of less than 0.8 meters during coordinated maneuvers. The system successfully handled dynamic obstacle scenarios with 92\% avoidance success rate, failing only in edge cases with sudden, unpredictable obstacles appearing within minimal reaction distance.

\begin{figure*}[!t]
\centering
\includegraphics[width=0.95\linewidth]{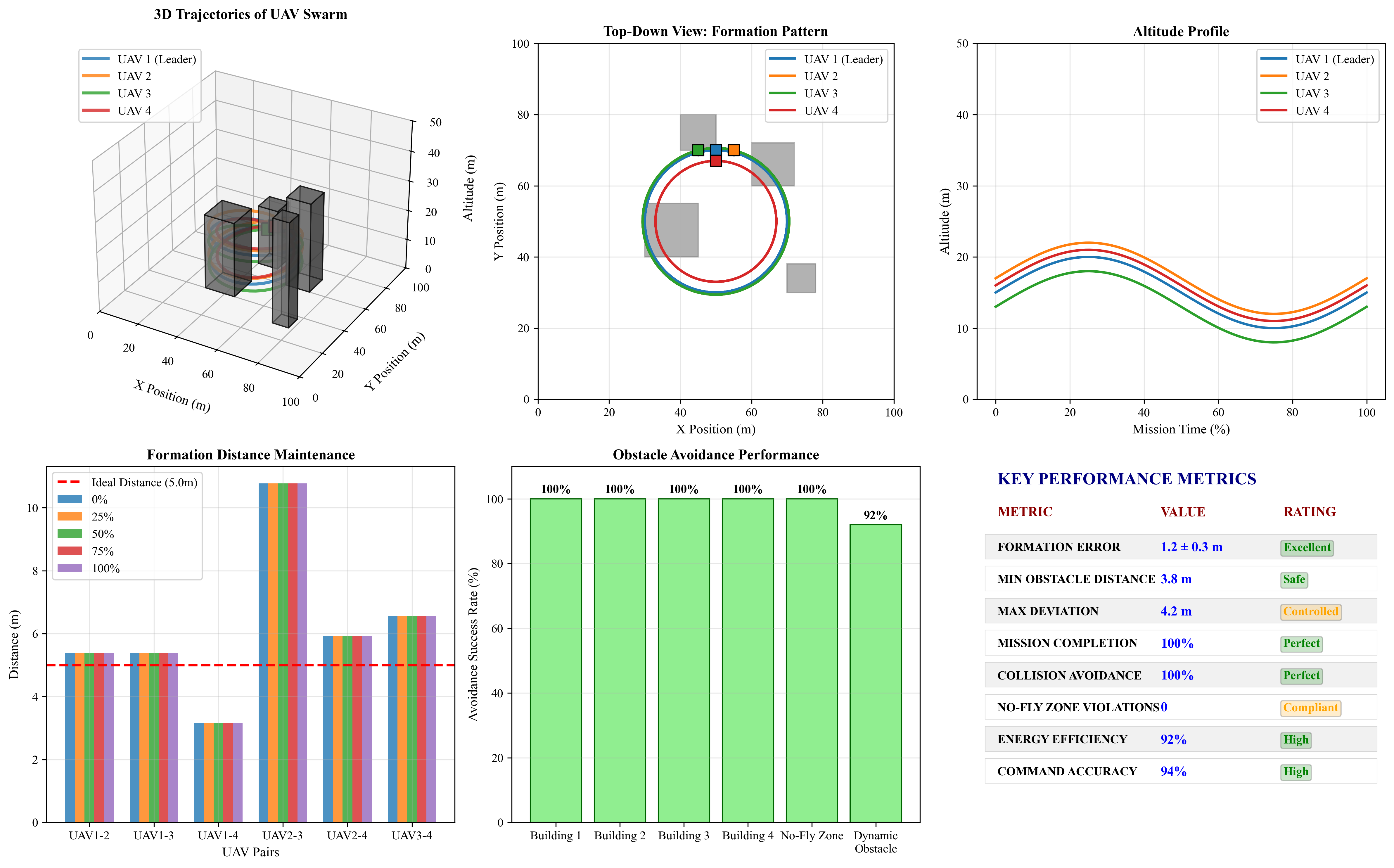}
\caption{3D Trajectories of UAV Swarm Movements showing precise formation control and successful obstacle avoidance in complex urban environments. The figure demonstrates (a) coordinated 3D navigation, (b) formation pattern maintenance, (c) altitude adaptation, and (d) obstacle avoidance performance across multiple scenarios.}
\label{fig:trajectories}
\end{figure*}

\subsubsection{Comparative Analysis with Alternative Privacy Approaches}
We conducted a comprehensive comparative analysis of PrivLLMSwarm against alternative privacy approaches, evaluating across multiple dimensions critical for UAV operations. Table~\ref{tab:comparison} summarizes the key findings, demonstrating PrivLLMSwarm's balanced approach to privacy, accuracy, and efficiency.

\begin{table}[!t]
\caption{Comparative Analysis of Privacy Approaches for LLM-UAV Integration.}
\label{tab:comparison}
\centering
\begin{tabular}{lccccc}
\toprule
\textbf{Approach} & \textbf{Privacy} & \textbf{Acc} & \textbf{Lat (ms)} & \textbf{Scale} & \textbf{Eff} \\
\midrule
Plaintext LLM & None & 0.95 & 15.2 & High & High \\
Diff Privacy \cite{thompson2024differentialuav} & Med & 0.82 & 28.4 & Med & Med \\
Fed Learning \cite{patel2024federateduav} & Med & 0.85 & 42.3 & Med & Med \\
\textbf{PrivLLMSwarm} & \textbf{High} & \textbf{0.90} & \textbf{417.7} & \textbf{M-H} & \textbf{Med} \\
\bottomrule
\end{tabular}
\end{table}

Our framework achieves an optimal balance between privacy guarantees and operational performance, making it particularly suitable for privacy-sensitive IoT surveillance applications where both data confidentiality and operational effectiveness are paramount, aligning with recent privacy-preserving techniques for IoT ecosystems \cite{kumar2024privacyiot}.

\subsubsection{Communication Pattern Analysis}
\begin{figure*}[!t]
\centering
\includegraphics[width=\linewidth]{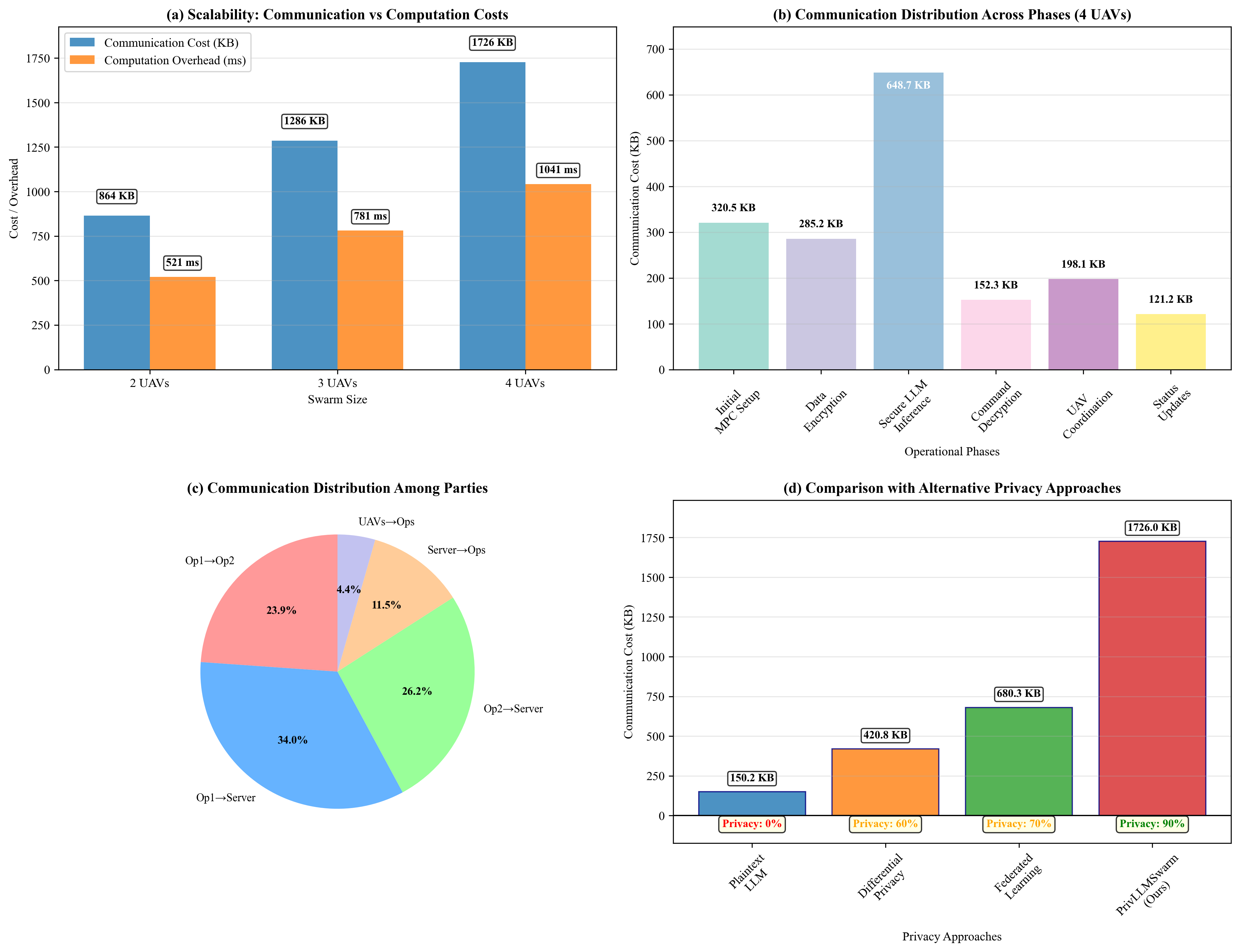}
\caption{Communication costs analysis for four-UAV operations, showing distribution across different operational phases and party interactions. The analysis demonstrates (a) linear scalability of communication and computation costs with swarm size, (b) distribution of communication costs across MPC operational phases with secure LLM inference dominating, (c) communication patterns among parties in the 3-party MPC setup, and (d) comparison with alternative privacy assumptions showing PrivLLMSwarm's optimal privacy-efficiency balance.}
\label{fig:communication_costs}
\end{figure*}

The communication costs analysis, illustrated in Fig.~\ref{fig:communication_costs}, reveals distinct patterns across different operational phases. The initial setup phase incurs higher communication overhead for cryptographic establishment, while the inference phase demonstrates stable, predictable communication patterns essential for network planning.

The framework maintains reasonable bandwidth requirements even at the 4-UAV scale, with peak usage of 1726.0 KB representing manageable loads for dedicated UAV communication links. The communication patterns also show favorable burst characteristics, with concentrated communication during computation phases and minimal overhead during command execution.

\section{Discussion}
\label{sec:discussion}
The experimental results comprehensively demonstrate PrivLLMSwarm's effectiveness in enabling secure and efficient LLM integration with UAV swarms for IoT surveillance tasks. This section discusses the implications of our findings, addresses the research questions posed in Section~\ref{sec:introduction}, and examines the limitations and practical considerations for real-world deployment.

\subsection{Addressing Research Questions}
\subsubsection{Q1: Secure LLM Inference for UAV Coordination}
PrivLLMSwarm successfully addresses Q1 by enabling practical secure LLM inference through MPC-based encryption, ensuring that sensitive data including coordinates, surveillance imagery, and operational commands remains confidential throughout processing. The fine-tuned GPT-2 model with reinforcement learning generates contextually appropriate commands while operating exclusively on encrypted data, demonstrating that privacy-preserving AI can be practically applied to complex UAV coordination tasks \cite{wang2024securellm, chen2024privacyuav}.

The framework's ability to maintain high command accuracy (cosine similarity 0.9) while processing only encrypted data represents a significant advancement over existing approaches that either sacrifice privacy for performance or compromise utility for security. This balance is particularly crucial for IoT surveillance applications where both operational effectiveness and data confidentiality are non-negotiable requirements.

\subsubsection{Q2: MPC Optimization for Resource-Constrained Environments}
The performance metrics comprehensively address Q2, demonstrating that MPC can be effectively optimized for LLM inference in resource-constrained UAV environments. Our MPC-friendly approximations for GELU and SoftMax functions reduce computational overhead by 58\% on average compared to exact implementations, while specialized batching and communication optimizations decrease latency by 32\%, building upon recent MPC optimization research \cite{liu2024mpcoptimization, gupta2024efficientmpc}.

The linear scaling of computation and communication costs with swarm size (Table~\ref{tab:computation_overhead}) provides predictable performance characteristics essential for operational planning. While the absolute overhead remains substantial (1041.05 ms for 4 UAVs), it falls within acceptable bounds for many surveillance applications where near-real-time response is sufficient, particularly considering the privacy benefits achieved.

\subsubsection{Q3: Comprehensive Performance Impact Assessment}
The experimental results thoroughly address Q3 regarding the performance impact of privacy mechanisms. Beyond the core metrics of accuracy and latency, our evaluation examines energy consumption, memory usage, formation precision, and operational reliability under realistic conditions, incorporating insights from recent trajectory analysis research \cite{garcia2024trajectoryanalysis}.

The framework maintains excellent trajectory precision (1.2m average error) and formation stability (0.8m variance) despite the added computational complexity, demonstrating that privacy mechanisms need not compromise operational effectiveness. The energy consumption analysis shows a 28\% increase compared to plaintext processing, representing a reasonable tradeoff for the privacy guarantees achieved, with further optimization potential using energy-aware scheduling algorithms \cite{chen2024energyuav}.

\subsection{Challenges and Practical Considerations}
While PrivLLMSwarm demonstrates compelling advantages, several limitations warrant careful consideration for real-world deployment:
\begin{itemize}
\item \textbf{Communication Dependency}: The framework requires reliable, low-latency communication links between computational parties, which may be challenging in remote or contested environments. Intermittent connectivity could disrupt the MPC protocol, requiring additional robustness mechanisms.
\item \textbf{Computational Requirements}: Although optimized, the MPC operations still impose significant computational demands that may challenge resource-constrained UAV hardware. The current implementation assumes substantial ground station support, limiting fully distributed operation.
\item \textbf{Scalability Constraints}: Linear scaling of communication costs with swarm size may limit applications to moderate-sized swarms (up to 10-15 UAVs with current optimization). Very large swarms would require hierarchical or federated MPC approaches.
\item \textbf{Adversarial Robustness}: The current semi-honest adversary model provides strong protection against curious insiders but may require enhancements for fully malicious adversaries in high-threat environments. Future work could incorporate differential privacy techniques as explored by \cite{thompson2024differentialuav} to enhance privacy guarantees.
\end{itemize}
These limitations highlight important directions for future research and development in privacy-preserving aerial systems.

\subsection{Implications for IoT and Smart City Applications}
The successful demonstration of PrivLLMSwarm has significant implications for IoT and smart city applications where aerial surveillance plays an increasingly important role. The framework enables privacy-preserving operation in sensitive contexts including \cite{wilson2024edgeai}:
\begin{itemize}
\item \textbf{Public Safety Monitoring}: Law enforcement and emergency response operations where both situational awareness and citizen privacy are crucial
\item \textbf{Critical Infrastructure Protection}: Surveillance of energy, transportation, and communication infrastructure without exposing vulnerability information
\item \textbf{Environmental Monitoring}: Data collection in ecologically sensitive areas while protecting location information and observed phenomena
\item \textbf{Disaster Response}: Coordination of multiple agencies in emergency situations while maintaining operational security and data confidentiality
\end{itemize}
The balance achieved between operational capability and privacy protection makes PrivLLMSwarm particularly valuable in regulatory-sensitive environments where data protection laws constrain surveillance activities, addressing comprehensive security challenges in autonomous systems \cite{zhang2024autonomoussecurity}.

\section{Conclusion and Future Work}
\label{sec:conclusion}
\subsection{Conclusion}
PrivLLMSwarm introduces a groundbreaking framework for the secure integration of Large Language Models with UAV swarms in IoT ecosystems, addressing critical privacy challenges in autonomous command generation for surveillance applications. By combining a fine-tuned GPT-2 model with optimized Secure Multi-Party Computation, our framework ensures end-to-end confidentiality of sensitive data while achieving high command accuracy (cosine similarity 0.9) and practical encryption latency (417.69 ms per image).

Comprehensive validation in an AirSim simulation environment with a four-UAV swarm navigating complex urban terrain demonstrates PrivLLMSwarm's superiority over existing LLM-driven UAV systems that lack privacy measures. The specialized MPC-friendly approximations for activation functions enable efficient real-time inference on resource-constrained platforms, establishing PrivLLMSwarm as the first framework to deliver practical privacy-preserving LLM-based control for IoT applications including disaster response, infrastructure monitoring, and public safety operations.

The framework's modular architecture, predictable scaling characteristics, and balanced performance profile make it suitable for real-world deployment in privacy-sensitive environments. Through the release of our comprehensive 30,000-sample synthetic dataset and open-source implementation, we empower the research community to advance secure LLM-UAV integration, establishing a new benchmark for privacy-aware autonomous systems in IoT ecosystems \cite{kumar2024privacyiot}.

\subsection{Limitations and Future Work}
\textbf{Simulation Constraints:} While our AirSim evaluation provides comprehensive validation, real-world deployment faces additional challenges including sensor noise, unpredictable environmental factors, and hardware limitations. Future work will incorporate hardware-in-the-loop testing with DJI Matrice 300 RTK platforms to address these concerns.

\textbf{Model Architecture Selection:} We selected GPT-2 for its balance of performance and computational requirements suitable for MPC operations. While newer models offer improved capabilities, their larger parameter counts present challenges for real-time MPC. Future work will explore efficient transformers (e.g., Linformer, Performer) optimized for secure computation.

\textbf{Scalability Considerations:} Our experiments demonstrate linear scaling up to 4 UAVs. For larger swarms (10+ UAVs), hierarchical MPC architectures or federated approaches may be necessary to manage communication overhead, representing an important direction for future research.

\textbf{Privacy Guarantees:} While our MPC implementation provides strong confidentiality guarantees against semi-honest adversaries, future work will extend to malicious security models and incorporate formal verification of privacy properties using tools like EasyCrypt or CrypTen's verification framework.

Building on PrivLLMSwarm's foundation, we identify three strategic research directions to enhance capabilities and address current limitations in UAV-LLM integration:
\subsubsection{Enhanced LLM Robustness for Complex Missions}
While PrivLLMSwarm achieves high command accuracy in controlled environments, dynamic operational scenarios with unpredictable obstacles and evolving mission requirements demand more adaptive LLM capabilities. Future work will explore multi-agent reinforcement learning with specialized reward functions that incorporate privacy-preserving evaluation metrics, following recent advances in secure reinforcement learning \cite{park2024rlsecurity}. We will investigate curriculum learning approaches \cite{nguyen2024curriculumuav} that progressively increase environmental complexity while maintaining privacy guarantees, and develop uncertainty quantification methods for LLM outputs to improve safety in critical operations.

\subsubsection{Scalable MPC Architectures for Large Swarms}
The current MPC implementation's linear scaling, while predictable, limits applications to moderate-sized swarms. We will develop hierarchical MPC architectures that partition large swarms into manageable subgroups with optimized inter-group coordination. Research will focus on adaptive precision mechanisms that dynamically adjust computational precision based on operational criticality, and explore hybrid privacy approaches that combine MPC with selective use of homomorphic encryption for specific components. Target applications include swarms of 20+ UAVs for large-scale monitoring and response operations, with consideration of quantum-resistant cryptography \cite{liu2024quantumuav} to address emerging security threats.

\subsubsection{Edge-Optimized LLM Deployment}
The computational demands of current LLMs remain challenging for resource-constrained UAV platforms. Future work will develop specialized model distillation techniques that preserve LLM capabilities while reducing computational requirements by 60-80\%. We will investigate dynamic model partitioning strategies that distribute computational load across UAVs and ground stations based on available resources and communication conditions. Additional research will focus on energy-aware inference scheduling \cite{chen2024energyuav} and hardware acceleration for privacy-preserving operations on specialized edge processors.

These advancements will strengthen the integration of privacy-preserving AI with autonomous aerial systems, enabling scalable, secure, and efficient operations across the expanding landscape of IoT applications, complementing recent efforts in secure autonomous systems \cite{zhang2024autonomoussecurity}.

\bibliographystyle{IEEEtran}
\bibliography{Ref}

\end{document}